\documentstyle[aps,multicol,prl,epsf]{revtex}
\begin{document}
\title{Measuring the Kondo effect in the Aharonov-Bohm interferometer}
\author{Amnon Aharony and Ora Entin-Wohlman}
\address{Department of Physics, Ben Gurion University, Beer
Sheva 84105, Israel, School of Physics and Astronomy, \\Tel Aviv
University, Tel Aviv 69978, Israel, and Argonne National
Laboratory, Argonne, IL 60439}
\date{\today}
\maketitle

\begin{abstract}
The conductance ${\cal G}$ of an Aharonov-Bohm interferometer
(ABI), with a strongly correlated quantum dot on one arm, is
expressed in terms of the dot Green function, $G_{dd}$, the
magnetic flux $\phi$ and the non-interacting parameters
 of the ABI. We show that
one can extract $G_{dd}$ from the observed oscillations of ${\cal
G}$ with $\phi$, for both closed and open ABI's. In the latter
case, the phase shift $\beta$ deduced from ${\cal G} \approx
A+B\cos(\phi+\beta)$ depends strongly on the ABI's parameters, and
 usually  $\beta \ne \pi/2$. These parameters may also
 reduce the Kondo temperature, eliminating the
Kondo behavior.
\\[3ex]
\end{abstract}


\begin{multicols}{2}

The recent observation of the Kondo effect in quantum dots (QD's),
whose parameters can be tuned continuously \cite{mit}, has been
followed by much theoretical 
and experimental activity. For temperatures $T$ below the Kondo
temperature $T_K$, the spin of an electron localized on the QD is
dynamically screened by the electrons in the Fermi sea, yielding a
large conductance ${\cal G}$ through the QD, close to the unitary
value $2e^2/h$, and a transmission phase $\alpha$ equal to $\pi/2$
\cite{langreth,oreg}. A good tool to test these predictions
involves embedding the strongly-correlated QD on one arm of an
Aharonov-Bohm interferometer (ABI). Indeed, such experiments were
carried out for both a closed (two-terminal) ABI \cite{vander} and
an open (multi-terminal) ABI \cite{ji}. Both experiments exhibited
the Aharonov-Bohm oscillations with the normalized flux
$\phi=e\Phi/\hbar c$. The former experiments exhibited the
expected ``phase rigidity", with ${\cal G}$ an even function of
$\phi$ \cite{rigid}. However, there has been no {\it quantitative}
analysis of these data. The latter experiments attempted to
measure the transmission phase, and found a variety of behaviors
which were inconsistent with the expected value of $\pi/2$. As a
result, Ji {\it et al.} \cite{ji} stated that ``the full
explanation of the Kondo effect may go beyond the framework of the
Anderson model". Theoretical attempts to discuss related issues
have concentrated on the dot alone (when it is detached from the
ABI) \cite{oreg,silvestrov}, or applied various techniques
\cite{hofstetter,bulka,david} to the QD on simple models of the
closed ABI. However, it has not been very clear how to make
quantitative comparisons of theory and experiment.

Most of the theoretical discussions of QD's concentrate on the
retarded Green function for electrons with energy $\omega$ on the
QD, $G_{dd}(\omega)$ (we ignore the spin index, since we assume no
magnetic asymmetry). For a simple QD, connected to a broad
electronic band, the $T=0$ transmission amplitude for electrons
going through the QD is proportional to $G_{dd}(\epsilon_F)$,
where $\epsilon_F$ is the Fermi energy (taken as zero below)
\cite{langreth,meir}. However, measuring the conductance only
yields $|G_{dd}(0)|^2$, with no information on the phase. The ABI
experiments were thus intended to measure {\it both} the magnitude
and the phase of $G_{dd}$, and compare with theory. In this paper
we concentrate on the following question: given experimental data
on the flux dependent conductance of the ABI, ${\cal G}(\phi)$,
how can we deduce the ``intrinsic" Green function $G_{dd}$? An
earlier paper \cite{prl} answered this question for
non-interacting electrons on a simple model for a closed ABI, and
made some speculations on the interacting case. Another paper
\cite{abi} showed that for non-interacting electrons, the phase
shift measured in the open ABI depends on details of the opening.
Here we show that for strongly correlated electrons, ${\cal
G}(\phi)$ is much more sensitive to the specific details of the
ABI's. We give explicit instructions for extracting $G_{dd}$ from
the measured ${\cal G}(\phi)$, and show that the opening has much
stronger effects in the Kondo regime, possibly explaining the
puzzling experiments \cite{ji}.

\begin{figure}
\leavevmode \epsfclipon \epsfxsize=9truecm
\hspace{-.5cm}\vbox{\epsfbox{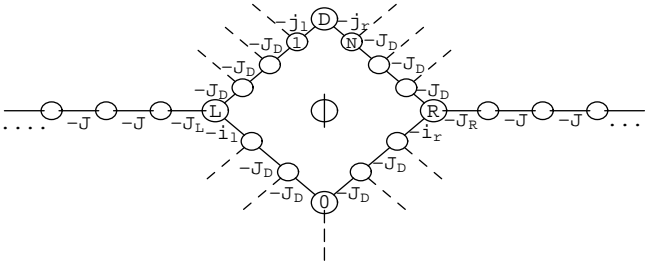}}
\vspace{0.2cm} \caption{The model for the ABI, with
$n_l=n_r=3,~n_0=5$.} \label{ABI}
\end{figure}

Our qualitative results should apply for a large class of ABI's.
For simplicity, we demonstrate them for the specific Anderson
model shown in Fig. \ref{ABI}, which captures the important
ingredients. The conductance ${\cal G}$ is measured between the
two leads which are attached to sites ``L" and ``R" on the ABI
ring. The QD (denoted ``D") is connected to L (R) via $n_l$
($n_r$) sites. The lower ``reference" branch contains $n_0$ sites.
Except for the QD, we use a tight binding model, with the real
hopping matrix elements as indicated in the figure. Site energies
are $\epsilon_l,~\epsilon_r$ and $\epsilon_0$ on the respective
branches, $\epsilon_L,~\epsilon_R$ on sites L and R and zero on
the leads. Using gauge invariance, we introduce the normalized
flux $\phi$ as a phase factor in $J_{D1}=J_{1D}^\ast=j_l
e^{i\phi}$. The Hamiltonian on the dot is
${\cal H}_d=\epsilon_d\sum_\sigma
n_{d\sigma}+Un_{d\uparrow}n_{d\downarrow}$,
with obvious notations. Here we assume that the transport is
dominated by the level $\epsilon_d$ on the QD. We also assume that
$U$ is very large, and ignore the resonance at $2\epsilon_d+U$.
Figure \ref{ABI} generalizes earlier models
\cite{hofstetter,bulka}, by adding the internal structure on the
links between D, L and R. Some such structure always exists in
experiments, and may have important effects on the observed
conductance (see below). For the open ABI, each dashed line
represents an additional lead, with a hopping matrix element
$-J_X$ on its first bond \cite{abi}.

For $T \ll T_K$, it is sufficient to calculate ${\cal G}$ at
$T=0$. Irrespective of the above details, one has
\begin{eqnarray}
{\cal G}=\frac{2e^2}{h}4\Gamma_L(0)\Gamma_R(0)\big
|G_{LR}(\omega=0)\big |^2,\label{cond1}
\end{eqnarray}
where $G_{LR}(\omega)$ is the Fourier transform of the retarded
Green function involving a creation of an electron at site R and
an annihilation of an electron at site L at time $t$ later
\cite{caroli}. Also,
$\Gamma_{L,R}(\omega)=-\Im\Sigma^0_{L,R}(\omega)$, where
$\Sigma^0_{L,R}$ is the self-energy generated at sites L or R due
to the leads.
Since interactions exist only on the dot, one can use the
equations of motion to express all the retarded Green functions
$G_{\alpha\beta}$ in terms of $G_{dd}$. If $g_{\alpha\beta}$
denotes the Green functions of the whole network without the QD
and the two bonds connected to it, then it is straightforward to
obtain \cite{kondo} the relation
\begin{eqnarray}
G_{LR}=g_{LR}+\sum_{\alpha\beta}g_{L\alpha}J_{\alpha
D}J_{D\beta}g_{\beta R} G_{dd}, \label{GLR}
\end{eqnarray}
where in our model the only non-zero $J_{\alpha\beta}$'s are
$J_{D1}=J_{1D}^\ast=j_l e^{i\phi}$ and $J_{DR}=J_{RD}=j_r$ (see
Fig. \ref{ABI}).

For the non-interacting case, one has \begin{eqnarray}
G_{dd}=1/[\omega-\epsilon_d-\Sigma_0(\omega)], \label{non}
\end{eqnarray}
with the non-interacting self-energy
\begin{eqnarray}
\Sigma_0(\omega)=
\sum_{\alpha\beta}J_{D\alpha}g_{\alpha\beta}(\omega)J_{\beta D}
\equiv \delta\epsilon_d(\omega)-i\Delta_0(\omega).
\end{eqnarray}
Since the phase $\phi$ is only contained in $J_{D1}=J_{1D}^\ast$,
the complex matrix $g_{\alpha\beta}$ is independent of $\phi$ and
obeys $g_{\alpha\beta}=g_{\beta\alpha}$. Therefore,
\begin{eqnarray}
\Sigma_0(\omega)=j_l^2 g_{11}+j_r^2 g_{NN}+2j_lj_r g_{1N}\cos\phi
\label{SIG0}
\end{eqnarray}
is an even function of $\phi$. Except for the very special case
$n_l=n_r=n_0=0$ (discussed e.g. in Refs. \cite{hofstetter,bulka}),
$g_{1N}$ is not a real number, and therefore {\it both}
$\delta\epsilon_d(\omega)$ {\it and} $\Delta_0(\omega)$ oscillate
with $\phi$,
\begin{eqnarray}
\delta\epsilon_d=a_1+b_1\cos\phi;\ \
\Delta_0=a_2+b_2\cos\phi.\label{delphi}
\end{eqnarray}

Using these expressions, Eq. (\ref{GLR}) can be written as
\begin{eqnarray}
G_{LR}=g_{LR}G_{dd}\Bigl
([G_{dd}]^{-1}+\Sigma_0+\sum_{\alpha\beta}
J_{D\alpha}w_{\alpha\beta}J_{\beta D}\Bigr ),
\end{eqnarray}
with
$w_{\alpha\beta}\equiv g_{\alpha
R}g_{L\beta}/g_{LR}-g_{\alpha\beta}$.
In calculating the necessary $g_{\alpha\beta}$'s, it is convenient
to first eliminate all the leads, replacing them by site
self-energies. For example, at site L we replace the bare Green
function $1/(\omega-\epsilon_L)$ by
$1/[\omega-\epsilon_L-\Sigma^0_L(\omega)]$. For our
one-dimensional leads one has
$\Sigma^0_L(\omega)=-e^{i|q|a}J_L^2/J$, where $\omega=-2J\cos(qa)$
is the energy of the electron in the band of the leads (we assume
all the leads to have the same $J$ and the same lattice constant
$a$) \cite{kondo}. The results remain valid also for more complex
leads, as long as one uses the appropriate band-generated
self-energies. For the remaining $N=n_l+n_r+n_0+2$ sites of the
ring (without the QD), $g_{\alpha\beta}$ is then the inverse of a
tri-diagonal $N \times N$ matrix,
${\cal M}_{\alpha\beta}=J_{\alpha\beta}+[\omega-\epsilon_\alpha-
\Sigma^0_\alpha(\omega)]\delta_{\alpha\beta}$.
For such a matrix, it turns out that one always has $w_{1N}=0$.
Thus,
\begin{eqnarray}
{\cal G}={\cal G}_{ref}|G_{dd}|^2\big |G_{dd}^{-1}+\Sigma_0+x+y
e^{-i\phi}\big |^2,\label{cond}
\end{eqnarray}
where $x=j_l^2 w_{11}+j_r^2 w_{NN}$, $y=j_l j_r w_{N1}$ and the
``reference" conductance ${\cal G}_{ref}\equiv (2e^2/h)
4\Gamma_L(0)\Gamma_R(0)|g_{LR}|^2$ {\it depend only on the
parameters of the non-interacting parts of the ABI, and not on the
QD parameters} $\epsilon_d$ and $U$.

At $T=0$ the electrons must obey the Fermi liquid relations
\cite{bickers1}. Specifically,  at the Fermi energy one expects
\begin{eqnarray} \Im[G_{dd}^{-1}] (\omega=0) \equiv
\Delta_0(\omega=0). \label{FL} \end{eqnarray} This condition
should hold for {\it any} network in which the QD is embedded, and
is therefore true for {\it both the closed and the open ABI}. In
the limit of a very large negative $\epsilon_d$, when $\langle
n_d\rangle \rightarrow 1$, one also expects that $\Re[G_{dd}^{-1}]
(\omega=0)\rightarrow 0$. For the simple QD with two leads, one
has $\Sigma_0(\omega)=\Sigma^0_L(\omega)+\Sigma^0_R(\omega)$, and
in the symmetric case $\Gamma_L=\Gamma_R=\Delta_0/2$ the
conductance reaches its unitary limit $2e^2/h$ [Eq. (\ref{cond1}),
with $G_{LR}\rightarrow G_{dd}$]. In this limit the phase of
$G_{dd}$ becomes $\pi/2$ \cite{langreth}.

\begin{figure}
\leavevmode \epsfclipon \epsfxsize=4.5truecm
\hspace{-.5cm}\vbox{\epsfbox{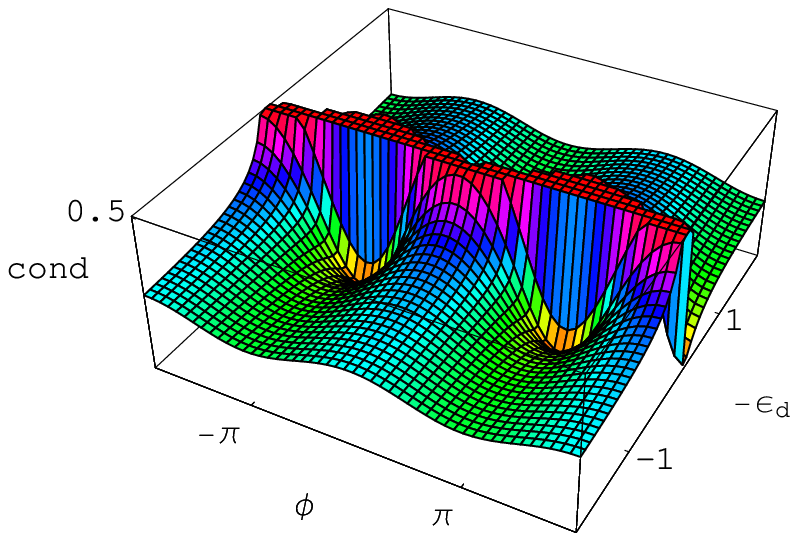}}{\leavevmode
\epsfclipon \epsfxsize=4.5truecm\hspace{0cm}\epsfbox{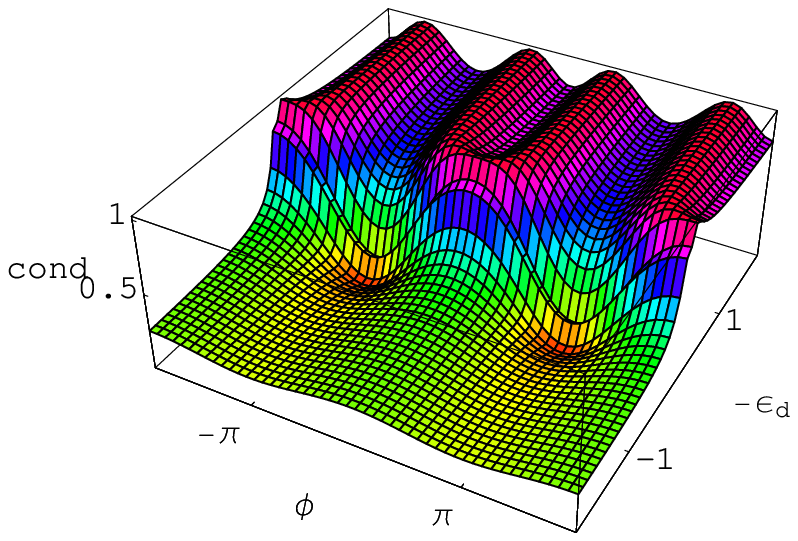}}
\vspace{.1cm}
%
\caption{Conductance (in units of $2e^2/h$) through the closed ABI
versus the normalized flux $\phi$ and the energy of the state on
the dot $\epsilon_d$ (the gate voltage), without (LHS) and with
interactions (RHS). $n_l=n_r=2,~n_0=3$, $J_L=J_R=J_D=1$,
$j_l=j_r=.2,~i_l=i_r=.4$,
$\epsilon_l=\epsilon_r=\epsilon_0=-.3,~\epsilon_L=\epsilon_R=0$.
All energies are in units of $J$.} \label{closed}
\end{figure}

We now discuss the closed ABI. In this case, $w_{11},~w_{NN}$ and
$w_{N1}$ and therefore also $x$ and $y$ are all real numbers.
Using the Fermi liquid result (\ref{FL}) and Eq. (\ref{delphi}),
Eq. (\ref{cond}) becomes
\begin{eqnarray}
\frac{{\cal G}_{closed}}{{\cal G}_{ref}}\equiv{\cal F}(\phi)\equiv
\frac{(\zeta+r_a+r_b\cos\phi)^2+
r_y^2\sin^2\phi}{\zeta^2+(1+r_d\cos\phi)^2},\label{GGG}
\end{eqnarray}
with the dimensionless function $\zeta(\phi)=\Re[G_{dd}^{-1}]/a_2$
and constants $r_a=(a_1+x)/a_2,~r_b=(b_1+y)/a_2, ~r_y=y/a_2$ and
$r_d=b_2/a_2$. For the non-interacting case,
$\zeta=[-\epsilon_d-\delta\epsilon_d(0)]/a_2$, and Eq. (\ref{GGG})
generalizes the results of Ref. \cite{prl}. An example is shown on
the LHS of Fig. \ref{closed}. In the strongly correlated case and
in the unitary limit, $\zeta=0$ and ${\cal G}_{closed} \rightarrow
{\cal G}_{ref}{\cal F}_0(\phi)$. All the features in the
$\phi$-dependence of ${\cal F}_0$ arise {\it only due to the
non-interacting parts of the ABI}. Usually, Eq. (\ref{GGG})
contains many harmonics. Except in special cases
\cite{hofstetter}, it is {\it not} dominated by the second
harmonic, and the period of ${\cal F}_0(\phi)$ is not simply
doubled. An example of this dependence is seen (for large negative
$\epsilon_d$) on the RHS of Fig. \ref{closed}: except for the
minima at $\phi=0$ and $\pi$, the maxima are {\it not} at $\pi/2$.
Experimentally, one knows that one has reached this limit once the
function ${\cal G}_{closed}(\phi)$ no longer changes with the gate
voltage which governs $\epsilon_d$. The reference conductance
${\cal G}_{ref}$ can be measured by disconnecting the QD, i.e.
setting $j_l=j_r=0$. Alternatively, ${\cal G}_{ref}$ can be
absorbed in the scales of the parameters in the numerator of Eq.
(\ref{GGG}). Having determined ${\cal G}_{ref}$, one can determine
the four real parameters $r_a,~r_b,~r_y$ and $r_d$ by a fit to
${\cal F}_0(\phi)$ (In practice, one only needs four values of the
function) \cite{foot}. Having found these parameters, one can now
move away from the unitary limit, and measure ${\cal
G}_{closed}={\cal G}_{ref}{\cal F}(\phi)$. The unknown function
$\zeta(\phi)$ can now be found from the quadratic equation
\begin{eqnarray}
\zeta^2-2\zeta\frac{r_a+r_b\cos\phi}{{\cal F}-1}+\frac{{\cal
F}-{\cal F}_0}{{\cal F}-1}(1+r_d\cos\phi)^2=0.
\end{eqnarray}
The solution should be chosen so that it decreases to zero at
large negative $\epsilon_d$ and increases linearly with large
positive $\epsilon_d$. Having found the solution, the phase
$\alpha$ of $G_{dd}$ is then defined via
\begin{eqnarray}
\cot\alpha=-\frac{\Re[G_{dd}^{-1}](\omega=0)}{\Delta_0(\omega=0)}
\equiv -\frac{\zeta}{1+r_d\cos\phi}.\label{beta}
\end{eqnarray}
This phase, or equivalently $\Re[G_{dd}^{-1}]$, are the quantities
obtained from theories.

For demonstrating the qualitative dependence of ${\cal G}$ on
$\phi$ and on the other parameters, we have used an approximate
analytic  solution of the equations of motion, truncated via
decoupling of higher order Green functions \cite{kondo}. In the
limits $T=\omega=0$ and $U\rightarrow \infty$, this solution
assumes the simple analytic form
\begin{eqnarray}
\cot\alpha=-\frac{z\delta n-\frac{3}{4}\Bigl (z+\sqrt{z^2+\delta
n(\frac{3}{2}-\delta n)}\Bigr )}{2\Bigl [\frac{3}{4}\delta
n+z\Bigl (z+\sqrt{z^2+\delta n(\frac{3}{2}-\delta n)}\Bigr )\Bigr
]},\label{EOM}
\end{eqnarray}
where $z$ represents the value at $\omega=0$ of the
non-interacting ratio
\begin{eqnarray}
z(\omega)=[\omega-\epsilon_d-\delta\epsilon_d(\omega)]/[2\Delta_0(\omega)],\label{zzz}
\end{eqnarray}
while $\delta n$ is related to the electron occupation on the dot
via $\langle n_d\rangle=2(1-\delta n)$ (which should be determined
self-consistently). In practice, $\delta n$ varies smoothly
between $1/2$ (at $z \gg 1$) and 1 (at $z \ll -1$), and the
results of calculations are not very sensitive to the details of
this variation. Equation (\ref{EOM}) interpolates between
$\beta=\pi/2$ at $z\rightarrow \infty$ and $\beta = \pi$ for $z
\rightarrow -\infty$. In the latter limit, $G_{dd}$ approaches the
{\it non-interacting} form (\ref{non}). Using Eq. (\ref{EOM}) in
Eq. (\ref{GGG}) for a specific set of parameters yields the RHS of
Fig. \ref{closed}. One clearly sees the transition from the
non-interacting behavior at large positive $\epsilon_d$ (compare
with the LHS) to the unitary limit at large negative $\epsilon_d$.
For different sets of parameters one reproduces qualitatively all
the earlier results, including the Fano-Kondo effect
\cite{hofstetter}. We have used these results to imitate real
experimental ``data", and were able to use the above algorithm to
extract $\cot\alpha$ as in Eq. (\ref{beta}).

Note that the above analysis yields $G_{dd}$ for the QD {\it on
the ABI}, where this function (and thus also the phase $\alpha$)
depends explicitly on the flux $\phi$, via $z$. At $T=\omega=0$,
we expect $\alpha$ to depend only on the ratio $z$ also for other
theories. In our case, $z$ can be extracted from the experimental
data via
\begin{eqnarray}
z=-[\tilde \epsilon_d+r_a+(r_b-r_y)\cos\phi]/(1+r_d\cos\phi),
\label{zz}
\end{eqnarray}
where $\tilde \epsilon_d=(\epsilon_d-x)/a_2$ is just a shifted
rescaled gate voltage. Having deduced the dependence of both $z$
and $\alpha$ on $\phi$, a parametric plot can yield $\alpha$
versus $z$, for comparison with single dot calculations.
Alternatively, one can experimentally study the results as
function of the coupling to the reference branch, $i_l$ and $i_r$.
Extrapolation to $i_l,~i_r \rightarrow 0$ would give the
dependence of $\alpha$ on $\Sigma_0(0)$ for the upper branch
alone. However, $\Sigma_0(\omega)$ still depends on the finite
chains connecting D with L and R \cite{pascal}.

\begin{figure}
\leavevmode \epsfclipon \epsfxsize=4.5truecm
\hspace{-.5cm}\vbox{\epsfbox{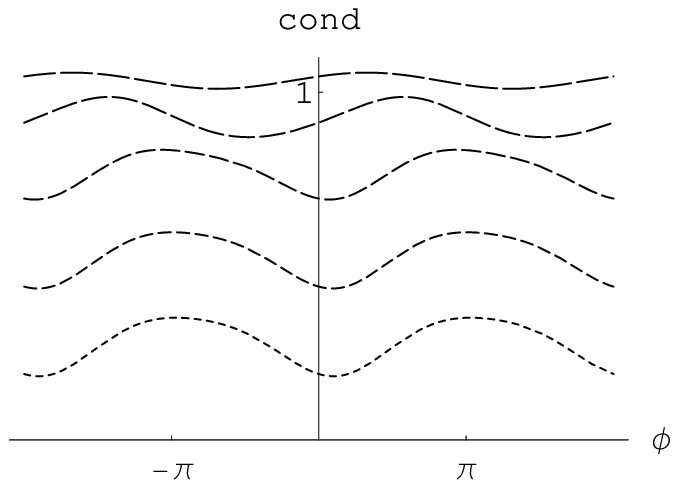}}{\leavevmode
\epsfclipon
\epsfxsize=4.5truecm\hspace{0cm}\epsfbox{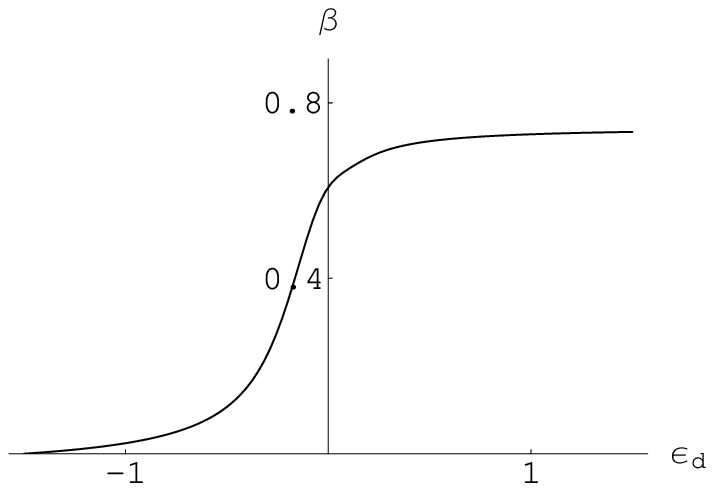}}
\vspace{.5 cm}\caption{LHS: Conductance through the open ABI
versus $\phi$, at $\epsilon_d=(-1.5,~-1,~-.5,~0,~.5,~1,~1.5)J$,
with interactions. Graphs are shifted up with increasing
$\epsilon_d$. Parameters are the same as in Fig. \ref{closed}, but
with $J_X=.5$. RHS: The ``measured" phase shift $\beta$ (in units
of $\pi$) versus $\epsilon_d$.} \label{open}
\end{figure}

We now turn to the open ABI, with $J_X \ne 0$. Equation
(\ref{cond})  remains correct, but now $x$ and $y$ become {\it
complex}. Interestingly, Eq. (\ref{SIG0}) still holds, and
$\Sigma_0$ is still an even function of $\phi$. In the unitary
limit,  ${\cal G}(\phi)$ has the exact form
\begin{eqnarray}
{\cal G}_{open} \rightarrow {\cal
G}_{ref}\frac{A+B\cos(\phi+\tilde\beta)+C\cos(2\phi+\gamma)}{(1+r_d\cos\phi)^2},
\label{Gopen}
\end{eqnarray}
and we need six parameters to fit it. Note that all the ABI
parameters (including ${\cal G}_{ref}$) now also depend on $J_X$.
The two lower curves in the left panel of Fig. \ref{open} show
results in this limit. Note that the graphs are not sinusoidal,
mainly due to the second term in the numerator and to the
denominator in Eq. (\ref{Gopen}). Since one remains close to the
Kondo resonance, the denominator continues to be important,
modifying the 2-slit-like numerator. The asymmetric shape of each
oscillation seems similar to that reported in Ref. \cite{ji}. The
other curves in the same panel were derived using Eq. (\ref{EOM}).
Again, one observes the crossover to the non-interacting
sinusoidal shape at large positive $\epsilon_d$. To extract a
``transmission phase" from these curves, one can e.g. follow the
maxima as function of $\epsilon_d$, or enforce a fit to the
two-slit formula ${\cal G}_{open} \approx A+B\cos(\phi+\beta)$.
Since now there is no well-defined zero to $\phi$, one can only
deduce the relative change in the phase $\beta$. Setting $\beta=0$
at $\epsilon_d\rightarrow -\infty$, the RHS of Fig. \ref{open}
shows this relative phase versus $\epsilon_d$. For the parameters
we used, the total change is about $0.8\pi$, far away from the
expected change in $\alpha$, equal to $\pi/2$. The actual values
depend on details of the ABI. This may explain the non-trivial
values of the phases observed in Ref. \cite{ji}: they result from
the experimental setup, and not from a breakdown of the Anderson
theory.

Finally, a few words about non-zero $T$ or $\omega$. Generally,
$T$ and $\omega$ enter into $G_{dd}$ similarly. In the approximate
solution of Ref. \cite{kondo}, one ends up with a competition
between the variable $z(\omega)$ of Eq. (\ref{zzz}) and
$\log(D/T)$ or $\log(D/\omega)$, where $2D=4J$ is the width of the
band in the leads. This competition yields estimates of $T_K$,
\begin{eqnarray}
\log(T_K/D) = \pi[\epsilon_d+\delta\epsilon_d(0)]/\Delta_0(0).
\end{eqnarray}
Although more accurate theories end up with different expressions,
all of them end up with a strong dependence on the ratio which
appears on the RHS. In our case, this ratio oscillates strongly
with $\phi$, opening the possibility that for different fluxes the
QD is below or above $T_K$. We emphasize the appearance of
$\delta\epsilon_d$ in the numerator,  ignored in some papers.

At non-zero $T$, the ``intrinsic" phase of the QD is expected to
start at 0 for large negative $\epsilon_d$ [where
$T>T_K(\epsilon_d)$], then grow to $\pi/2$ for intermediate
negative $\epsilon_d$'s (the unitary region), and finally grow to
$\pi$ at positive $\epsilon_d$ \cite{oreg}. As mentioned, both
$\delta\epsilon_d$ and $\Delta_0$ depend on the opening parameter
$J_X$. Using the approximation of Ref. \cite{kondo} also for
$T>0$, we found that large values of $J_X$ may completely
eliminate the intermediate plateau in $\alpha$, and give a direct
increase of $\alpha$ from 0 to $\pi$. Unlike the non-interacting
case \cite{abi}, where changing $J_X$ only slightly modified the
quantitative shape of the function $\beta(\epsilon_d)$, the
effects here are {\it qualitative}: opening may lower $T_K$ and
completely eliminate the observability of the Kondo behavior.
Again, this could have happened in Ref. \cite{ji}.

We acknowledge helpful discussions with Y. Imry, Y. Meir, P. Simon
and A. Schiller. This project was carried out in a center of
excellence supported by the ISF under grant No. 1566/04. Work at
Argonne supported by the U. S. Department of Energy, Basic Energy
Sciences--Materials Sciences, under Contract \#W-31-109-ENG-38.

\end{multicols}

\begin{references}
\bibitem{foot} In fact, the fit to Eq. (\ref{GGG})
only determines the combinations $(r_a\pm r_b)^2$ and
$r_a^2+r_y^2$, which allows four different fits. This ambiguity
can be removed by going to large positive $\epsilon_d$, where
$G_{dd}$ approaches Eq. (\ref{non}),
$\zeta=[-\epsilon_d-\delta\epsilon_d(0)]/a_2$ and one can follow
the variation of ${\cal G}_{closed}$ with $\epsilon_d$.
\end{references}
\end{document}